\documentclass[submission,copyright,creativecommons]{eptcs}
\providecommand{\event}{TARK 2023}

\usepackage{underscore}
\usepackage[T1]{fontenc}
\usepackage{microtype}

\title{Exploiting Asymmetry in Logic Puzzles: Using ZDDs for Symbolic Model Checking Dynamic Epistemic Logic}

\author{
  Daniel Miedema
  \institute{Bernoulli Institute \\ University of Groningen \\ The Netherlands}
  \email{daniel2Miedema@gmail.com}
  \and
  Malvin Gattinger
  \institute{ILLC \\ University of Amsterdam \\ The Netherlands}
  \email{malvin@w4eg.eu}
}

\def\titlerunning{Exploiting Asymmetry in Logic Puzzles: ZDDs for Symbolic Model Checking DEL}
\def\authorrunning{D.~Miedema, M.~Gattinger}

\newcommand{\negInv}{\raisebox{\depth}{\scalebox{1}[-1]{$\neg$}}}
\newcommand{\Then}{{\normalfont\textsc{Then}}}
\newcommand{\Else}{{\normalfont\textsc{Else}}}
\newcommand{\flipLeaf}{\mathsf{flipLeaf}}
\newcommand{\flipEdge}{\mathsf{flipEdge}}

\usepackage{amsmath}
\usepackage{amssymb}
\usepackage{float}
\usepackage{caption}
\usepackage{color}
\usepackage{mathtools}
\usepackage{multicol}
\usepackage{graphicx}
\usepackage{subcaption}
\usepackage{xparse}
\usepackage{verbatim}

\usepackage{amsthm}
\newtheorem{theorem}{Theorem}
\newtheorem{definition}[theorem]{Definition}
\newtheorem{fact}[theorem]{Fact}
\newtheorem{example}[theorem]{Example}

\usepackage{pgfplots}
\pgfplotsset{compat=1.17}
\usepackage{pgfplotstable}
\usetikzlibrary{plotmarks}
\usepgfplotslibrary{external}
\definecolor{BDD}{rgb}{0.54901,0.33725,0.29411}  
\definecolor{BDDc}{rgb}{0.17254,0.62745,0.17254} 
\definecolor{T0}{rgb}{0.12156,0.46666,0.70588}   
\definecolor{E0}{rgb}{0.58039,0.40392,0.74117}   
\definecolor{T1}{rgb}{1.0,0.49803,0.054901960}   
\definecolor{E1}{rgb}{0.83921,0.15294,0.15686}   

\usepackage{listings}
\begin{document}

\maketitle

\begin{abstract}
  Binary decision diagrams (BDDs) are widely used to mitigate the state-explosion problem in model checking.
  A variation of BDDs are Zero-suppressed Decision Diagrams (ZDDs) which omit variables that must be false, instead of omitting variables that do not matter.

  We use ZDDs to symbolically encode Kripke models used in Dynamic Epistemic Logic, a framework to reason about knowledge and information dynamics in multi-agent systems.
  We compare the memory usage of different ZDD variants for three well-known examples from the literature: the Muddy Children, the Sum and Product puzzle and the Dining Cryptographers.
  Our implementation is based on the existing model checker SMCDEL and the CUDD library.

  Our results show that replacing BDDs with the right variant of ZDDs can significantly reduce memory usage.
  This suggests that ZDDs are a useful tool for model checking multi-agent systems.
\end{abstract}

\section{Introduction}\label{sec:introduction}

There are several formal frameworks for reasoning about knowledge in multi-agent systems, and many are implemented in the form of epistemic model checkers.
Here we are concerned with the \emph{data structures} used in automated epistemic reasoning.
This is a non-issue in theoretical work, where Kripke models are an elegant mathematical tools.
But they are not very efficient: models where agents know little tend to be the largest.
More efficient representations are often based on Binary Decision Diagrams (BDDs), which use the idea that a representation of a function not depending on $p$ can simply ignore that variable $p$.
This fits nicely to the models encountered in epistemic scenarios, such as the famous example of the Muddy Children: If child $2$ does not observe whether it is muddy, i.e.\ whether $p_2$ is true or false, then we can save memory by omitting $p_2$ in the encoding of the knowledge of child $2$.
However, which variables matter may change, and in many examples the claim that ``many variables do not matter'' only holds in the initial model.
This motivates us to look at Zero-suppressed Decision Diagrams (ZDDs) which use an asymmetric reduction rule to omit variables that \emph{must} be \emph{false}, instead of the symmetric reduction rule targeting variables that \emph{do not matter}.

Our informal research question is thus: Is it more memory efficient to have a default assumption that ``anything we do not mention does not matter'' or, for example ``anything we do not mention must be false''?
Obviously, the answer will depend on many aspects.
Here we make the question precise for the case of Dynamic Epistemic Logic, and consider three well-known examples from the literature.

The article is structured as follows.
We discuss related work in the rest of this section, then we provide the relevant background in Sections~\ref{sec:theory-dds} and~\ref{sec:theory-del}.
Section~\ref{sec:methods} describes our experiment design and the formal models used.
We present our results in Section~\ref{sec:results} and conclude in Section~\ref{sec:conclusion}.

\paragraph{Related work}

Model checking aims to verify properties of formally specified systems.
Standard model checking methods search through a whole state transition graph and thus suffer from the state explosion problem: the number of states grows exponentially with the number of components or agents.
To tackle this problem \emph{symbolic} methods were developed~\cite{burch1992symbolic}.
These reduce the amount of resources needed, by reasoning about sets instead of individual states.
Starting with SMV from~\cite{mcmillan1993symbolic}, most approaches use Binary Decision Diagrams (BDDs)~\cite{bryant1986graph} to encode Boolean functions.
Zero-suppressed Decision Diagrams (ZDDs) are an adaption of BDDs, introduced by Minato~\cite{minato1993zero}.
ZDDs naturally fit combinatorial problems and many comparisons between BDDs and ZDDs have been done.
For both an elegant introduction into the topic of BDDs and many more references we refer to~\cite{knuth2011art4Ap1}.
Symbolic model checking using ZDDs has not been studied much, partly due to underdeveloped construction methods~\cite{minato2001zero}.

Most existing symbolic model checkers use temporal logics such as LTL or CTL\@.
Yet problems come in many forms and for examples typically described using epistemic operators (e.g.\ in multi-agent systems), Dynamic Epistemic Logic (DEL) is an established framework~\cite{van2007dynamic}.
Also DEL model checking can be done symbolically~\cite{vBvEGS2018:SMCDELbeyondS5}, by encoding Kripke models as so-called knowledge structures. This lead to its implementation, SMCDEL, which is extended in this work.
Another encoding, sometimes also called ``symbolic models'', is based on mental programs~\cite{charrierSymbolic2019}.
In concrete applications such as ``Hintikka's World'' these also get encoded as BDDs~\cite{charrierHintikkas2019}.
To our knowledge no previous work used ZDDs or other BDD variants for DEL model checking, with the exception of~\cite{gamblinSymbolic2022} where Algebraic Decision Diagrams (ADDs) are used for probabilistic DEL\@.

Here our main research questions is:
Can ZDDs be more compact than BDDs when encoding the Kripke models for classical logic puzzles?
We answer this question by adding ZDD functionality to SMCDEL and then comparing the sizes for three well-known examples from the literature.

\section{Theory: Decision Diagrams}\label{sec:theory-dds}

Symbolic model checkers, including SMCDEL, rely on efficient representations of Boolean functions.
The most widely used data structure for this are Binary Decision Diagrams (BDDs).
In this section we recall their definition and explain the difference between standard BDDs and ZDDs.
How Boolean functions are then used for model checking DEL will be explained in the next section.
Before we get to decision diagrams we define Boolean formulas and functions.

\begin{definition}
  The \emph{Boolean formulas} over a set of variables $P$ (also called \emph{vocabulary}) are given by
  $\varphi ::= \top \mid p \mid \lnot \varphi \mid \varphi \land \varphi$
  where $p \in P$.
  We define
  $\bot := \lnot \top$,
  $\varphi \lor \psi := \lnot (\lnot \varphi \land \lnot \psi)$
  and
  $\varphi \rightarrow \psi := \lnot (\varphi \land \lnot \psi)$.

  We write $\vDash$ for the usual Boolean semantics using assignments of type $P \to \{0, 1\}$.
  When $P$ is given we identify an assignment (also called \emph{state}) with the set of variables it maps to $1$.
  A \emph{Boolean function} is any $f \colon \mathcal{P}(P) \to \{0,1\}$.
  For any $\varphi$ we define the Boolean function
  $f_\varphi(s) := \{ \text{if } s \vDash \varphi \text{ then } 1 \text{ else } 0 \}$.
\end{definition}

For example, if our vocabulary is $P = \{p,q,r\}$ and  $s(p) = 0$, $s(q)=1$ and $s(r)=0$ then we identify $s$ with $\{q\}$ and we have $s \vDash (\lnot p \land q) \lor r$.
In the following we will also just write $\varphi$ for $f_\varphi$.
Notably, two different formulas can correspond to the same Boolean function, but not vice versa.

\begin{definition}\label{def:SubstitQuantif}
  For any $\varphi$, $\psi$, and $p$,
  let $\varphi(\frac{p}{\psi})$ be the result of replacing every occurrence of $p$ in $\varphi$ by ${\psi}$.
  For any $A=\{p_1,\dots,p_n\}$,
  let $\varphi(\frac{A}{\psi}) := \psi (\frac{p_1}{\psi}) (\frac{p_2}{\psi}) \dots (\frac{p_n}{\psi}) $.
  We use $\forall p \varphi $ to denote $\varphi\left(\frac{p}\top\right) \wedge \varphi\left(\frac{p}\bot\right)$.
  For any $A=\{p_1,\dots,p_n\}$, let $\forall A \varphi := \forall p_1 \forall p_2 \dots \forall p_n \varphi$.
\end{definition}

\paragraph{Decision Diagrams}

A decision diagram is a rooted directed acyclic graph, used to encode a Boolean function.
Any terminal node (i.e.\ leaf) is labelled with 0 or 1, corresponding to the result of the function.
Any internal node $n$ is labelled with a variable and has two outgoing edges to successors denoted by $\Then(n)$ and $\Else(n)$ --- each representing a possible value for the variable.
A path from the root to a leaf in a decision diagram corresponds to an evaluation of the encoded function.
A decision diagram is called \emph{ordered} if the variables are encountered in the same order on all its paths.

\begin{example}
  The first (left-most) decision diagram in Figure~\ref{fig:dd-sdd-bdd-zdd} is a full decision tree for $q \land \lnot r$.
  To evaluate it at state $\{p,q\}$ we start at the root and then go along the solid $\Then$-edge because $p$ is true, then again along a $\Then$-edge as $q$ is true and then along the dashed $\Else$-edge as $r$ is false.
  We get $1$ as a result, reflecting the fact that $\{p,q\} \vDash q \land \lnot r$.
  Similarly we can use the second and third diagram.

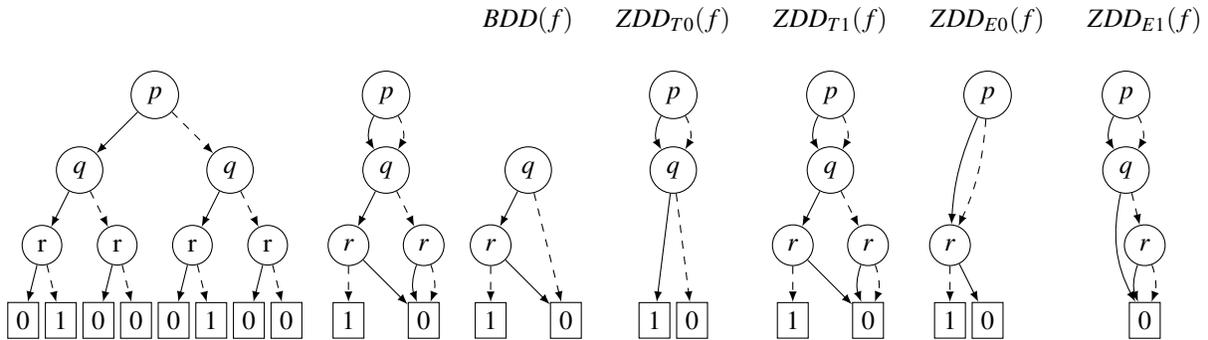
\begin{figure}[H]
  \centering
  \small
  \begin{tikzpicture}[->, >=latex]
    \node (root) at (5,10) [draw,circle] {$p$};
    \node (t) at (4,9) [draw, circle] {$q$};
    \node (f) at (6,9) [draw, circle] {$q$};
    \node (tt) at (3.5,8) [draw,circle] {r};
    \node (tf) at (4.5,8) [draw,circle] {r};
    \node (ft) at (5.5,8) [draw,circle] {r};
    \node (ff) at (6.5,8) [draw,circle] {r};
    \node (ttt) at (3.25,7) [draw,rectangle] {0};
    \node (ttf) at (3.75,7) [draw,rectangle] {1};
    \node (tft) at (4.25,7) [draw,rectangle] {0};
    \node (tff) at (4.75,7) [draw,rectangle] {0};
    \node (ftt) at (5.25,7) [draw,rectangle] {0};
    \node (ftf) at (5.75,7) [draw,rectangle] {1};
    \node (fft) at (6.25,7) [draw,rectangle] {0};
    \node (fff) at (6.75,7) [draw,rectangle] {0};
    \draw (root) -- (t);
    \draw (root) [dashed] -- (f);
    \draw (t) -- (tt);
    \draw (t) [dashed] -- (tf);
    \draw (f) -- (ft);
    \draw (f) [dashed] -- (ff);
    \draw (tt) -- (ttt);
    \draw (tt) [dashed] -- (ttf);
    \draw (tf) -- (tft);
    \draw (tf) [dashed] -- (tff);
    \draw (ft) -- (ftt);
    \draw (ft) [dashed] -- (ftf);
    \draw (ff) -- (fft);
    \draw (ff) [dashed] -- (fff);
  \end{tikzpicture}
  \hfill
  \begin{tikzpicture}[->, >=latex]
    \node (root) at (4,10) [draw,circle] {$p$};
    \node (t) at (4,9) [draw, circle] {$q$};
    \node (tt) at (3.5,8) [draw,circle] {$r$};
    \node (tf) at (4.5,8) [draw,circle] {$r$};
    \node (0) at (4.5,7) [draw,rectangle] {0};
    \node (1) at (3.5,7) [draw,rectangle] {1};
    \draw (root)          to [bend right=30] (t);
    \draw (root) [dashed] to [bend left=30] (t);
    \draw (t) -- (tt);
    \draw (t) [dashed] -- (tf);
    \draw (tt) -- (0);
    \draw (tt) [dashed] -- (1);
    \draw (tf)          to [bend right=20] (0);
    \draw (tf) [dashed] to [bend left=20] (0);
  \end{tikzpicture}
  \hfill
  \begin{tikzpicture}[->, >=latex]
    \node (label) at (4,11) {$BDD(f)$};
    \node (t) at (4,9) [draw, circle] {$q$};
    \node (tt) at (3.5,8) [draw,circle] {$r$};
    \node (0) at (4.5,7) [draw,rectangle] {0};
    \node (1) at (3.5,7) [draw,rectangle] {1};
    \draw (t) -- (tt);
    \draw (t) [dashed] -- (0);
    \draw (tt) -- (0);
    \draw (tt) [dashed] -- (1);
  \end{tikzpicture}
  \hfill
  \begin{tikzpicture}[->, >=latex]
    \node (label) at (4,11) {$ZDD_{T0}(f)$};
    \node (root) at (4,10) [draw,circle] {$p$};
    \node (t) at (4,9) [draw, circle] {$q$};
    \node (0) at (4.25,7) [draw,rectangle] {0};
    \node (1) at (3.75,7) [draw,rectangle] {1};
    \draw (root)          to [bend right=30] (t);
    \draw (root) [dashed] to [bend left=30] (t);
    \draw (t) -- (1);
    \draw (t) [dashed] -- (0);
  \end{tikzpicture}
  \hfill
  \begin{tikzpicture}[->, >=latex]
    \node (label) at (4,11) {$ZDD_{T1}(f)$};
    \node (root) at (4,10) [draw,circle] {$p$};
    \node (t) at (4,9) [draw, circle] {$q$};
    \node (tt) at (3.5,8) [draw,circle] {$r$};
    \node (tf) at (4.5,8) [draw,circle] {$r$};
    \node (0) at (4.5,7) [draw,rectangle] {0};
    \node (1) at (3.5,7) [draw,rectangle] {1};
    \draw (root)          to [bend right=30] (t);
    \draw (root) [dashed] to [bend left=30] (t);
    \draw (t) -- (tt);
    \draw (t) [dashed] -- (tf);
    \draw (tt) -- (0);
    \draw (tt) [dashed] -- (1);
    \draw (tf)          to [bend right=20] (0);
    \draw (tf) [dashed] to [bend left=20] (0);
  \end{tikzpicture}
  \hfill
  \begin{tikzpicture}[->, >=latex]
    \node (label) at (4,11) {$ZDD_{E0}(f)$};
    \node (root) at (4,10) [draw,circle] {$p$};
    \node (tt) at (3.5,8) [draw,circle] {$r$};
    \node (0) at (4,7) [draw,rectangle] {0};
    \node (1) at (3.5,7) [draw,rectangle] {1};
    \draw (root)          to [bend right=10] (tt);
    \draw (root) [dashed] to [bend left=10] (tt);
    \draw (tt) -- (0);
    \draw (tt) [dashed] -- (1);
  \end{tikzpicture}
  \hfill
  \begin{tikzpicture}[->, >=latex]
    \node (label) at (4.25,11) {$ZDD_{E1}(f)$};
    \node (root) at (4,10) [draw,circle] {$p$};
    \node (t) at (4,9) [draw, circle] {$q$};
    \node (tf) at (4.25,8) [draw,circle] {$r$};
    \node (0) at (4.25,7) [draw,rectangle] {0};
    \draw (root)          to [bend right=30] (t);
    \draw (root) [dashed] to [bend left=30] (t);
    \draw (t) to [bend right=20] (0);
    \draw (t) [dashed] -- (tf);
    \draw (tf)          to [bend right=20] (0);
    \draw (tf) [dashed] to [bend left=20] (0);
  \end{tikzpicture}
  \caption{Seven decision diagrams for $f := q \land \lnot r$, assuming vocabulary $\{p,q,r\}$.}\label{fig:dd-sdd-bdd-zdd}
\end{figure}
\end{example}

\emph{Binary Decision Diagrams} (BDDs) were introduced by~\cite{bryant1986graph} and are particularly compact decision diagrams, obtained using two reduction rules.
The first rule identifies isomorphic subgraphs, i.e.\ we merge nodes that have the same label and the same children.
In Figure~\ref{fig:dd-sdd-bdd-zdd} we get from the first to the second diagram.
The second rule eliminates redundant nodes.
A node is considered redundant if both its $\Then$- and $\Else$-edge go to the same child.
In Figure~\ref{fig:dd-sdd-bdd-zdd} this gets us from the second to the third diagram.

\emph{Zero-suppressed Decision Diagrams} (ZDDs) were introduced by~\cite{minato1993zero} and use a different second rule than BDDs.
While in BDDs a node $n$ is eliminated when $\Then(n) = \Else(n)$, in ZDDs a node is eliminated when $\Then(n) = 0$.
In Figure~\ref{fig:dd-sdd-bdd-zdd} this rule gets us from the second to the fourth diagram called $ZDD_{T0}(f)$.
The idea is to not ignore the variables that ``do not matter'' (as $p$ in $q \land \lnot r$), but to remove the nodes of variables that must be false (as $r$ in $q \land \lnot r$).
To evaluate $ZDD_{T0}(f)$ at state $\{p,q\}$ we again start at the root and twice follow a solid edge because $p$ and $q$ are true, but then we notice that the solid edge goes from $q$ to $1$, without asking for the remaining variable $r$.
When evaluating a $ZDD_{T0}$ such a transition demands that the variable we ``jump over'' must be false --- hence the name ``zero-suppressed''.
Indeed $r$ is false in our state, so we do reach $1$.
If $r$ would have been true, the result would have been $0$.

\paragraph{Generalizing Elimination Rules}
The elimination rule ``remove nodes that have a $\Then$-edge leading to $0$'' can be modified in two obvious ways: instead of $\Then$- we could consider $\Else$-edges, and instead of $0$ we could consider $1$.
This leads us to three additional elimination rules.
\begin{definition}\label{def:elimRules}
  We denote five different node elimination rules as follows.
  A node $n$ with pairs of children $(\Then(n), \Else(n))$ is eliminated if it matches the left side of the rule,
  and any edges leading to $n$ are diverted to the successor $s$ on the right side of the rule.
  \[
    \begin{array}{l@{\hspace{4em}}l@{\hspace{4em}}l}
      EQ: \ \ (s, s) \Rightarrow s & T0: \ \ (0, s) \Rightarrow s & E0: \ \ (s, 0) \Rightarrow s\\
                                   & T1: \ \ (1, s) \Rightarrow s & E1: \ \ (s, 1) \Rightarrow s \\
    \end{array}
  \]
\end{definition}
Here $EQ$ is the rule for BDDs, while $T0$ (for ``Then $0$'') is the traditional ZDD rule.
The remaining three are variations.
For example, $E0$ says that any node with an $\Else$-edge to $0$ is removed, and any edge that led to the removed node should be diverted to where the $\Then$-edge of the removed node led.

In Figure~\ref{fig:dd-sdd-bdd-zdd} the $E0$ rule gets us from the second to the sixth diagram $ZDD_{E0}(f)$.
Note that we used the rule twice: After deleting an $r$ node the $q$ node has an $\Else$-branch to $0$, so it is also eliminated.
All diagrams encode the same function $f$, but when evaluating them we must interpret ``jumps'' differently.

A crucial feature of BDDs and ZDDs is that they are \emph{canonical} representations: given a fixed variable order there is a unique BDD and a unique ZDD for each variant.
It also becomes clear that for different Boolean functions a different kind of diagram can be more or less compact.

\begin{definition}
  For any Boolean function $f$, recall that $\neg f$ denotes its complement.
  Let $\negInv f$ denote the result of complementing all atomic propositions inside $f$.
  (For example, $\negInv (q \land \lnot r) = \lnot q \land r$.)
  For any decision diagram $d$,
  let $\flipLeaf(d)$ be the result of changing the labels of all leaves from 0 to 1 and vice versa;
  and let $\flipEdge(d)$ be the result of changing the labels of all edges from $\Then$ to $\Else$ and vice versa.
\end{definition}

There is a correspondence between $\neg$ and $\flipLeaf$, and between $\negInv$ and $\flipEdge$.
Moreover, we can use these operations to relate the four different variants of ZDDs as follows.

\begin{fact}\label{fct:genElimRules}
  For any Boolean function $f$ we have:
  \[ \begin{array}{lcl}
    DD_{T1}(f) & = & \flipLeaf \: DD_{T0} (\neg f) \\
    DD_{E0}(f) & = & \flipEdge \: DD_{T0} (\negInv f) \\
    DD_{E1}(f) & = & \flipEdge \: \flipLeaf \: DD_{T0} (\neg \negInv f)
  \end{array} \]
\end{fact}

\begin{example}
  We illustrate Fact~\ref{fct:genElimRules} using our running example $f := q \land \lnot r$ with vocabulary $\{p,q,r\}$.
  Figure~\ref{fig:zdd-var} shows the $T0$ decision diagrams mentioned in Fact~\ref{fct:genElimRules}.
  We see that for example $DD_{T1}(f)$ shown in Figure~\ref{fig:dd-sdd-bdd-zdd} is the same graph as $DD_{T0}(\neg f)$ with only the labels of the leaf nodes exchanged.
  Similarly, $DD_{E1}(f)$ in Figure~\ref{fig:dd-sdd-bdd-zdd} is the same graph as $DD_{T0}(\neg \negInv f)$ with flipped edges and leaves.
  \begin{figure}[H]
    \centering
    \begin{tikzpicture}[->,>=latex]
      \node (label) at (4,11) {$ZDD_{T0}(\neg f)$};
      \node (root) at (4,10) [draw,circle] {$p$};
      \node (t) at (4,9) [draw, circle] {$q$};
      \node (tt) at (3.5,8) [draw,circle] {$r$};
      \node (tf) at (4.5,8) [draw,circle] {$r$};
      \node (1) at (4.5,7) [draw,rectangle] {1};
      \node (0) at (3.5,7) [draw,rectangle] {0};
      \draw (root)          to [bend right=30] (t);
      \draw (root) [dashed] to [bend left=30] (t);
      \draw (t) -- (tt);
      \draw (t) [dashed] -- (tf);
      \draw (tt) -- (1);
      \draw (tt) [dashed] -- (0);
      \draw (tf)          to [bend right=20] (1);
      \draw (tf) [dashed] to [bend left=20] (1);
    \end{tikzpicture}
    \hspace{1em}
    \begin{tikzpicture}[->,>=latex]
      \node (label) at (4,11) {$ZDD_{T0}(\negInv f)$};
      \node (root) at (4,10) [draw,circle] {$p$};
      \node (tt) at (3.5,8) [draw,circle] {$r$};
      \node (0) at (4,7) [draw,rectangle] {0};
      \node (1) at (3.5,7) [draw,rectangle] {1};
      \draw (root) [dashed] to [bend right=10] (tt);
      \draw (root)          to [bend left=10] (tt);
      \draw (tt) [dashed] -- (0);
      \draw (tt)          -- (1);
    \end{tikzpicture}
    \hspace{0.5em}
    \begin{tikzpicture}[->,>=latex]
      \node (label) at (4,11) {$ZDD_{T0}(\neg \negInv f)$};
      \node (root) at (4,10) [draw,circle] {$p$};
      \node (t) at (4,9) [draw, circle] {$q$};
      \node (tf) at (4.25,8) [draw,circle] {$r$};
      \node (1) at (4.25,7) [draw,rectangle] {1};
      \draw (root) [dashed] to [bend right=30] (t);
      \draw (root)          to [bend left=30] (t);
      \draw (t) [dashed] to [bend right=20] (1);
      \draw (t)          to (tf);
      \draw (tf) [dashed] to [bend right=20] (1);
      \draw (tf)          to [bend left=20] (1);
    \end{tikzpicture}
    \caption{ZDDs with the same shape as the variants for $f := p \land \lnot q$.}\label{fig:zdd-var}
  \end{figure}
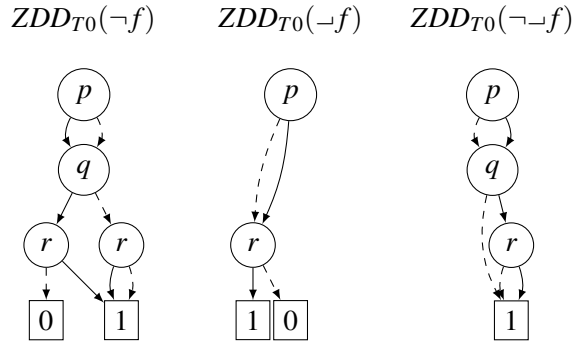
\end{example}

Fact~\ref{fct:genElimRules} is crucial for our implementation, because the CUDD library we use does not support $T1$, $E0$ and $E1$ explicitly.
Hence instead we always work with $T0$ diagrams of the negated or flipped functions.

\section{Theory: Symbolic Model Checking DEL}\label{sec:theory-del}

\paragraph{Kripke Models}\label{sec-kripke}
We recap the standard syntax and semantics of Public Announcement Logic (PAL), the most basic version of Dynamic Epistemic Logic (DEL).

\begin{definition}\label{def-syntax}
  Fix a vocabulary $V$ and a finite set of agents $I$.
  The DEL language $\mathcal{L}(V)$ is given by
  $\varphi ::= p \mid \lnot \varphi \mid \varphi \land \varphi
    \mid K_i \varphi
    \mid [\varphi]\varphi$
  where $p \in V$, $i \in I$.
\end{definition}

As usual, $K_i\varphi$ is read as \emph{``agent $i$ knows that $\varphi$''}.
The formula $[\psi]\varphi$ says that after a \emph{public announcement} of $\psi$, $\varphi$ holds.
The standard semantics for $\mathcal{L}(V)$ on Kripke models are as follows.

\begin{definition}\label{def-kr-models}
  A \emph{Kripke model} for a set of agents $I=\{1,\dots,n\}$ is a tuple
  ${\cal M}=(W,\pi ,{\cal K}_1 ,\dots,{\cal K}_n )$, where
  $W$ is a set of \emph{worlds},
  $\pi$ associates with each world a state $\pi(w)$,
  and
  ${\cal K}_1,\dots, {\cal K}_n $ are equivalence relations on $W$.
  A \emph{pointed Kripke model} is a pair $({\cal M}, w)$ consisting of a model and a world $w \in W$.
\end{definition}

\begin{definition}\label{def-kr-sat}
  Semantics for $\mathcal{L}(V)$ on pointed Kripke models are given inductively as follows.
  \begin{itemize}
  \item $({\cal M},w)\vDash p$ iff $\pi ^M (w)(p) = \top $.
  \item $({\cal M},w)\vDash \neg \varphi$ iff not $ ({\cal M},w)\vDash \varphi$
  \item $({\cal M},w)\vDash \varphi \land \psi$ iff $ ({\cal M},w)\vDash \varphi$ and $ ({\cal M},w)\vDash \psi$
  \item $ ({\cal M},w)\vDash {K}_i \varphi $ iff for all $w'\in W$,
    if $w{\cal K}_i^M w'$, then $({\cal M},w')\vDash  \varphi$.
  \item $ ({\cal M},w)\vDash [\psi] \varphi $ iff
    $({\cal M},w)\vDash \psi$ implies $({\cal M}^\psi,w)\vDash \varphi $
    where ${\cal M}^\psi$ is a new model based on the set $W^{{\cal M}^\psi} := \{ w \in W^{\cal M} \mid ({\cal M},w)\vDash \psi \}$
    and appropriate restrictions of ${\cal K}_i$ and $\pi$ to $W^{{\cal M}^\psi}$.
  \end{itemize}
\end{definition}

More expressive versions of DEL also include common knowledge and complex epistemic or ontic actions such as private communication, interception, spying and factual change.
Moreover, DEL can work both with S5 models and with arbitrary Kripke models.
All of this is compatible with the symbolic semantics we recall in the next section, but for our purposes in this article the restricted language above is sufficient, and we only consider S5 models.

\paragraph{Knowledge Structures}

While the semantics described above is standard, it has the disadvantage that models are represented explicitly, i.e.\ the number of worlds also determines the amount of memory needed to represent a model.
To combat this well-known state-explosion problem we can replace Kripke models with symbolic knowledge structures.
Their main advantage is that knowledge and results of announcements can be computed via purely Boolean operations, as shown in~\cite{vBvEGS2018:SMCDELbeyondS5}.

\begin{definition}\label{def-kns}
  Suppose we have $n$ agents.
  A \emph{knowledge structure} is a tuple
  $\mathcal{F} = (V,\theta,O_1,\dots,O_n)$ where
  $V$ is a finite set of atomic variables,
  $\theta$ is a Boolean formula over $V$ and
  for each agent $i$, $O_i\subseteq V$.
  The set $V$ is the \emph{vocabulary} and the formula $\theta$ is the \emph{state law} of $\mathcal{F}$.
  The $O_i$ are called \emph{observational variables}.
  An assignment over $V$ that satisfies $\theta$ is a \emph{state of $\mathcal{F}$}.
  A \emph{scene} is a pair $(\mathcal{F},s)$ where $s$ is a state of $\mathcal{F}$.
\end{definition}

\begin{example}
  Consider the knowledge structure $\mathcal{F} := ( V=\{p, q\}, \theta = {p \rightarrow q}, O_1=\{p\}, O_2=\{q\} )$.
  The states of $\cal F$ are the three assignments $\varnothing$, $\{q\}$ and $\{p,q\}$.
  Moreover, $\mathcal{F}$ has two agents who each observe one of the propositions: agent $1$ knows whether $p$ is true and agent $2$ knows whether $q$ is true.
\end{example}

We now give semantics for $\mathcal{L}(V)$ on knowledge structures.

\begin{definition}\label{def-kns-sat}
  Semantics for ${\cal L}(V)$ on scenes are defined as follows.
  \begin{itemize}
  \item $({\cal F},s)\vDash p$ iff $ s\vDash p$.
  \item $(\mathcal{F},s)\vDash \neg \varphi$ iff not $(\mathcal{F},s)\vDash \varphi$
  \item $(\mathcal{F},s)\vDash \varphi \land \psi$ iff $(\mathcal{F},s)\vDash \varphi$ and $(\mathcal{F},s)\vDash \psi$
  \item $(\mathcal{F},s) \vDash K_i \varphi$ iff
    for all $t$ of $\mathcal{F}$, if $s\cap O_i=t\cap O_i $, then $(\mathcal{F},t)\vDash \varphi$.
  \item $(\mathcal{F},s)\vDash [\psi] \varphi$ iff $(\mathcal{F},s)\vDash \psi$ implies $(\mathcal{F}^\psi, s) \vDash \varphi$
    where
    $\mathcal{F}^\psi:=(V,\theta \land \| \psi \|_\mathcal{F}, O_1, \dots, O_n) $.
  \end{itemize}
  where $\| \cdot \|_\mathcal{F}$ is defined in parallel in the following definition.
\end{definition}
\begin{definition}\label{def-boolEquiv}
  For any knowledge structure $\mathcal{F} = (V, \theta, O_1, \dots , O_n)$ and any formula $\varphi$ we define its \emph{local Boolean translation} $\| \varphi \|_\mathcal{F}$ as follows.
  \[
    \begin{array}{lcl@{\hspace{1.5cm}}lcl}
      \| p \|_\mathcal{F}                   & := & p                           & \| K_i \psi \|_\mathcal{F}     & := & \forall(V \setminus O_i)(\theta \rightarrow \| \psi \|_\mathcal{F}) \\
      \| \neg \psi \|_\mathcal{F}           & := & \neg \| \psi \|_\mathcal{F} & \| [ \psi ] \xi \|_\mathcal{F} & := & \| \psi \|_\mathcal{F} \rightarrow \| \xi \|_{\mathcal{F}^\psi}     \\
      \| \psi_1 \land \psi_2 \|_\mathcal{F} & := & \| \psi_1\|_\mathcal{F} \land \| \psi_2 \|_\mathcal{F}                                                                                  \\
    \end{array}
  \]
  where the case for $K_i \psi$ quantifies over the variables not observed by agent $i$,
  using Boolean quantification as defined in Definition~\ref{def:SubstitQuantif} above.
\end{definition}
A main result from~\cite{vBvEGS2018:SMCDELbeyondS5} based on~\cite{SuVarForget2009} is that for any finite Kripke model there is an equivalent knowledge structure and vice versa.
This means we can see knowledge structures as just another, hopefully more memory-efficient, data structure to store a Kripke model.
An additional twist is that we usually store the state law $\theta$ not as a formula but only the corresponding Boolean function --- which can be represented using a decision diagram as discussed in Section~\ref{sec:theory-dds}.

\section{Methods: Logic Puzzles as Benchmarks}\label{sec:methods}

Our leading question is whether ZDDs provide a more compact encoding than BDDs for models encountered in epistemic model checking.
To answer it we will work with three logic puzzles from the literature.
All examples start with an initial model which we encode as a knowledge structure with the state law as a decision diagram.
Then we make updates in the form of public announcements, changing the state law.
We record the size of the decision diagrams for each update step.

As a basis for our implementation and experiments we use \emph{SMCDEL}, the symbolic model checker for DEL from~\cite{vBvEGS2018:SMCDELbeyondS5}.
SMCDEL normally uses the BDD library CacBDD~\cite{lv2013cacbdd} which does not support ZDDs.
Hence we also use the library CUDD~\cite{somenzi2004cudd} which does support ZDDs.
However, also CUDD does not support the generalized elimination rules from Definition~\ref{def:elimRules}.
Therefore we use Fact~\ref{fct:genElimRules} to simulate the $T1$, $E0$ and $E1$ variants.
Our new code --- now merged into SMCDEL --- provides easy ways to create and update knowledge structures where the state law is represented using any of the four ZDD variants.

An additional detail is that CUDD always uses so-called complement edges to optimize BDDs, but not for ZDDs.
To compare the sizes of ZDDs to BDDs without complement edges we still use CacBDD\@.
Altogether in our data set we thus record the sizes of six decision diagrams for each state law: the EQ rule with and without complement edges (called BDD and BDDc) and the four ZDD variants from Definition~\ref{def:elimRules}.
We stress that by size of a diagram we mean the node count and not memory in bytes, because the former is independent of what libraries are used, whereas the latter depends on additional optimisations.

It now remains to choose examples.
We picked three well-known logic puzzles from the literature with different kinds of state laws, such that we also expect the advantage of ZDDs to vary between them.

\paragraph{Muddy Children}
The Muddy Children are probably the best-known example in epistemic reasoning, hence we skip the explanation here and refer to the literature starting with~\cite{littlewood1953methuen}.
A formalisation of the puzzle can be found in~\cite[Section~4.10]{van2007dynamic} and the symbolic encoding in~\cite[Section~4]{vBvEGS2018:SMCDELbeyondS5}.

\paragraph{Dining Cryptographers}
This problem and the protocol to solve it was first presented by~\cite{chaum1988dining}:
\begin{quote}
  ``Three cryptographers gather around a table for dinner. The waiter informs them that the meal has been paid for by someone, who could be one of the cryptographers or the National Security Agency (NSA). The cryptographers respect each other's right to make an anonymous payment, but want to find out whether the NSA paid.''
\end{quote}
The solution uses random coin flips under the table, each observed by two neighbouring cryptographers but not visible to the third one.
A formalisation and solution using Kripke models can be found in~\cite{van2007epistemic}.
To encode the problem in a knowledge structure we let $p_0$ mean that the NSA paid, $p_i$ for $i \in \{1,2,3\}$ that $i$ paid.
Moreover, $p_k$ for $k \in \{4,5,6\}$ represents a coin.
The initial scenario is then
$(V = \{p_0,\ldots,p_6\}, \theta = \otimes_1 \{p_0,p_1,p_2,p_3\}, O_1 = \{p_1,p_4,p_5\}, O_2 = \{p_2,p_4,p_6\}, O_3 = \{p_3,p_5,p_6\} )$
where the state law $\theta$ says that exactly one cryptographer or the NSA must have paid.
In the solution then each cryptographer announces the XOR ($\otimes$) of all bits they observe, with the exception that the payer should invert their publicly announced bit.
Formally, we get a sequence of three public announcements
$[?!(\otimes{p_1,p_4,p_5})] [?!(\otimes{p_2,p_4,p_6})] [?!(\otimes{p_3,p_5,p_6})]$
where $[?!\psi] \varphi := [!\psi] \varphi \land [\neg !\psi] \varphi$ abbreviates announcing whether.
The protocol can be generalised to any odd number $n$ instead of three participants.

\paragraph{Sum and Product}
The following puzzle was originally introduced in 1969 by H.~Freudenthal.
The translation is from~\cite{van2009publieke} where the puzzle is also formalised in DEL\@:
\begin{quote}
  A says to S and P\@:
  I have chosen two integers $x,y$ such that $1 < x < y$ and $x+y \leq 100$.
  In a moment, I will inform S only of $s = x + y$, and P only of $p = xy$.
  These announcements remain private.
  You are required to determine the pair $(x, y)$.
  He acts as said.
  The following conversation now takes place:
  P says: “I do not know it.” ---
  S says: “I knew you didn’t.” ---
  P says: “I now know it.” ---
  S says: “I now also know it.” ---
  Determine the pair (x, y).
\end{quote}

Solving the puzzle using explicit model checking is discussed in~\cite{vDRv2007::SAPinDEL}.
To represent the four variables and their values in propositional logic we need a binary encoding, using $\lceil \log_2 N \rceil$ propositions for each variable that take values up to $N$.
For example, to represent $x \leq 100$ we use $p_1,\ldots,p_7$ and encode the statement $x = 5$ as $p_1 \land p_2 \land p_3 \land p_4 \land \neg p_5 \land p_6 \land \neg p_7$, corresponding to the bit-string $0000101$ for $5$.

The initial state law for Sum and Product is a big disjunction over all possible pairs of $x$ and $y$ with the given restrictions, and the observational variables ensure that agents $S$ and $P$ know the values of $s$ and $p$ respectively.
For a detailed definition of the knowledge structure, see~\cite[Section~5]{vBvEGS2018:SMCDELbeyondS5}.

The announcements in the dialogue are formalised as follows, combining the first two into one:
First $S$ says $K_S \neg \bigvee_{i + j \leq 100} K_P ( x=i \land y=j )$, then $P$ says
  $\bigvee_{i + j \leq 100} K_P ( x=i \land y=j )$ and finally $S$ says
  $\bigvee_{i + j \leq 100} K_S ( x=i \land y=j )$.
Solutions to the puzzle are states where these three formulas can be truthfully announced after each other.
A common variation on the problem is to change the upper bound for $x+y$.
We use this to turn obtain a scalable benchmark, starting with 65 to ensure there exists at least one answer.

\bigskip

It is well known that ZDDs perform better on sparse sets~\cite{BryantBDDchapter}.
In our case, sparsity is the number of states in the model divided by the total number of possible states for the given vocabulary.
Our three examples vary a lot in their sparsity:
Muddy Children's sparsity is $0.5$ on average (going from $0.875$ to $0.125$, for 3 agents),
Dining Cryptographers is fairly sparse from start to finish ($0.25$ to $0.0625$, for 3 agents),
and Sum and Product is extremely sparse (e.g.\ starting with $< 1.369\cdot10^{-7}$ for $x+y \leq 100$).

\section{Results}\label{sec:results}

For each example we present a selection of results we deem most interesting, showing differences between BDD and ZDD sizes.
The full data set for two examples can be found in the appendix where we also include instructions how all of the results can be reproduced.

\paragraph{Muddy children}
We vary the number of children $n$ from 5 to 40, in steps of 5.
We can also vary the number of muddy children $m \leq n$, but mostly report results here where $m=n$.
Given any number of children, we record the size of the decision diagrams of the state law after the $k$th announcement, where $k$ ranges from $0$ (no announcements made yet) to $m-1$ (after which all children know their own state).

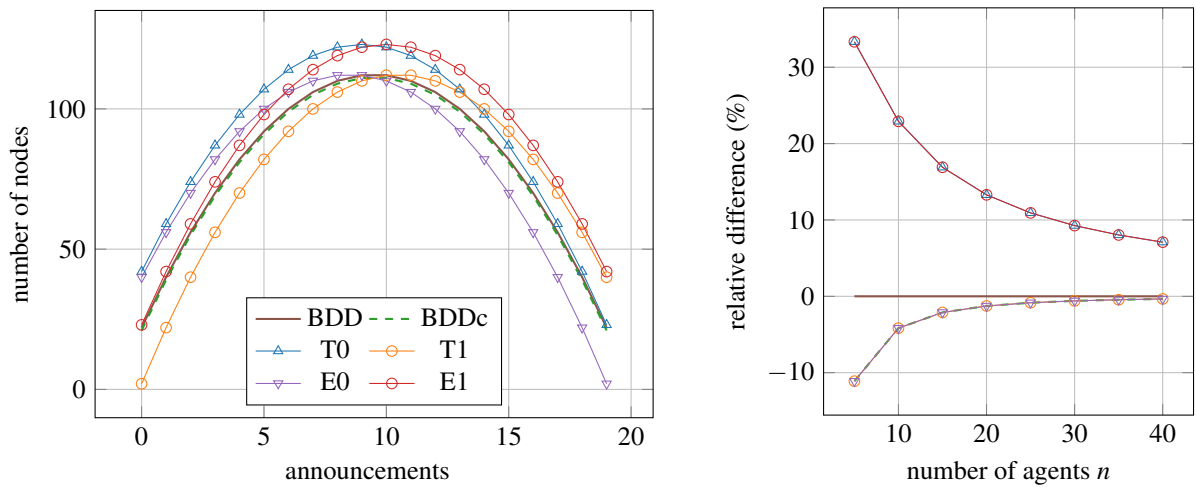
\begin{figure}[H]
  \centering
  \small
  \begin{subfigure}[b]{0.59\textwidth}
    \begin{tikzpicture}
      \pgfplotstableread{experimentResults/mc.dat}\mc
      \begin{axis}[legend columns=2, legend style={at={(0.5,0.03)},anchor=south}, width=9cm, height=7cm, grid=major,
        ylabel=number of nodes,
        xlabel=announcements\phantom{g},
        restrict expr to domain={\thisrow{round}}{0:19},
        restrict expr to domain={\thisrow{n}}{20:20},
        restrict expr to domain={\thisrow{m}}{20:20}]
        \addlegendentry{BDD};  \addplot+ [color=BDD, mark=none, thick] table[x=round, y=BDD] {\mc};
        \addlegendentry{BDDc}; \addplot+ [color=BDDc,mark=none, thick, dashed] table[x=round, y=BDDc] {\mc};
        \addlegendentry{T0};   \addplot+ [color=T0,  mark=triangle] table[x=round, y=T0] {\mc};
        \addlegendentry{T1};   \addplot+ [color=T1,  mark=o, solid] table[x=round, y=T1] {\mc};
        \addlegendentry{E0};   \addplot+ [color=E0,  mark=triangle, mark options={rotate=180}] table[x=round, y=E0] {\mc};
        \addlegendentry{E1};   \addplot+ [color=E1,  mark=o, solid] table[x=round, y=E1] {\mc};
      \end{axis}
    \end{tikzpicture}
    \caption{Absolute sizes per announcement, for $n=20$.}\label{fig:MC-resultplot}
  \end{subfigure}
  \begin{subfigure}[b]{0.4\textwidth}
    \begin{tikzpicture}[solid]
      \pgfplotstableread{experimentResults/mc.dat}\mc
      \begin{axis}[width=6.5cm, height=7cm, grid=major,
        ylabel=relative difference (\%),
        xlabel=number of agents $n$,
        restrict expr to domain={\thisrow{round}}{-1:-1},
        filter discard warning=false, unbounded coords=discard, x filter/.expression={\thisrow{n}==\thisrow{m} ? x : nan}]
        \addplot+ [color=BDD, mark=none, thick] table[x=n, y expr=0] {\mc};
        \addplot+ [color=BDDc,mark=none, thick, dashed] table[x=n, y expr=(\thisrow{BDDc}-\thisrow{BDD}) / \thisrow{BDD}*100] {\mc};
        \addplot+ [color=T0, mark=triangle] table[x=n, y expr=(\thisrow{T0}-\thisrow{BDD}) / \thisrow{BDD}*100] {\mc};
        \addplot+ [color=T1, mark=o, solid] table[x=n, y expr=(\thisrow{T1}-\thisrow{BDD}) / \thisrow{BDD}*100] {\mc};
        \addplot+ [color=E0, mark=triangle, mark options={rotate=180}] table[x=n, y expr=(\thisrow{E0}-\thisrow{BDD}) / \thisrow{BDD}*100] {\mc};
        \addplot+ [color=E1, mark=o, solid] table[x=n, y expr=(\thisrow{E1}-\thisrow{BDD}) / \thisrow{BDD}*100] {\mc};
      \end{axis}
    \end{tikzpicture}
    \caption{Relative average sizes.\phantom{p}}\label{fig:mc-relative}
  \end{subfigure}
  \caption{Results for Muddy Children.}
\end{figure}

As an example, let us fix $n=m=20$.
Figure~\ref{fig:MC-resultplot} shows the size of the decision diagrams after each announcement.
The lines all follow a similar curve, with the largest relative differences in the initial and final states.
Initially the most compact variant is T1 whereas at the end E0 is the most compact.
This matches the asymmetry in the Muddy Children story: at the start the state law is $p_1 \lor\ldots\lor p_n$, hence all $\Then$ edges lead to $1$ and $T1$ removes all nodes.
In contrast, at the end the state law is $p_1 \land\ldots\land p_n$ which means that all $\Else$ edges lead to $0$ and thus $E0$ eliminates all nodes.

Hence at different stages different variants are more compact.
But we want a representation that is compact throughout the whole process.
We thus consider the average size over all announcements, varying $n$ from $5$ to $40$.
Figure~\ref{fig:mc-relative} shows the relative size differences, with standard BDDs as 100\%.
The $T0$/$E1$ and the BDDc/$E0$/$T1$ lines overlap.
We see that $T1$ and $E0$ are more compact for small models, but not better than BDDs with complement edges and this advantage shrinks with a larger number of agents.

We also computed sizes for $m < n$, i.e.\ not all children being muddy.
In this case the sizes for each update step stay the same but there are fewer update steps because the last truthful announcement is in round $m-1$.
As expected this is in favour of the $T1$ variant.

\begin{figure}[H]
  \centering
  \small
  \begin{subfigure}[b]{0.59\textwidth}
    \begin{tikzpicture}
      \pgfplotstableread{experimentResults/dining.dat}\dining
      \begin{axis}[legend columns=2, legend pos=north west, width=9cm, height=7cm, grid=major,
        scaled ticks=false, tick label style={/pgf/number format/fixed},
        ylabel=number of nodes,
        xlabel=announcements\phantom{g},
        restrict expr to domain={\thisrow{round}}{0:},
        restrict expr to domain={\thisrow{n}}{13:13}]
        \addlegendentry{BDD};  \addplot+ [color=BDD, mark=none, thick] table[x=round, y=BDD] {\dining};
        \addlegendentry{BDDc}; \addplot+ [color=BDDc,mark=none, thick, dashed] table[x=round, y=BDDc] {\dining};
        \addlegendentry{T0};   \addplot+ [color=T0,  mark=triangle] table[x=round, y=T0] {\dining};
        \addlegendentry{T1};   \addplot+ [color=T1,  mark=o, solid] table[x=round, y=T1] {\dining};
        \addlegendentry{E0};   \addplot+ [color=E0,  mark=triangle, mark options={rotate=180}] table[x=round, y=E0] {\dining};
        \addlegendentry{E1};   \addplot+ [color=E1,  mark=o, solid] table[x=round, y=E1] {\dining};
      \end{axis}
    \end{tikzpicture}
    \caption{Absolute sizes per announcement, for $n=13$.}\label{fig:dining-resultplot}
  \end{subfigure}
  \begin{subfigure}[b]{0.4\textwidth}
    \begin{tikzpicture}[solid]
      \pgfplotstableread{experimentResults/dining.dat}\dining
      \begin{axis}[width=6.5cm, height=7cm, grid=major,
        ylabel=relative difference (\%),
        xlabel=number of agents $n$,
        restrict expr to domain={\thisrow{round}}{-1:-1},
        filter discard warning=false, unbounded coords=discard]
        \addplot+ [color=BDD, mark=none, thick] table[x=n, y expr=0] {\dining};
        \addplot+ [color=BDDc,mark=none, thick, dashed] table[x=n, y expr=(\thisrow{BDDc}-\thisrow{BDD}) / \thisrow{BDD}*100] {\dining};
        \addplot+ [color=T0, mark=triangle] table[x=n, y expr=(\thisrow{T0}-\thisrow{BDD}) / \thisrow{BDD}*100] {\dining};
        \addplot+ [color=T1, mark=o, solid] table[x=n, y expr=(\thisrow{T1}-\thisrow{BDD}) / \thisrow{BDD}*100] {\dining};
        \addplot+ [color=E0, mark=triangle, mark options={rotate=180}] table[x=n, y expr=(\thisrow{E0}-\thisrow{BDD}) / \thisrow{BDD}*100] {\dining};
        \addplot+ [color=E1, mark=o, solid] table[x=n, y expr=(\thisrow{E1}-\thisrow{BDD}) / \thisrow{BDD}*100] {\dining};
      \end{axis}
    \end{tikzpicture}
    \caption{Relative average sizes.\phantom{p}}\label{fig:dining-relative}
  \end{subfigure}
  \caption{Results for Dining Cryptographers.}
\end{figure}
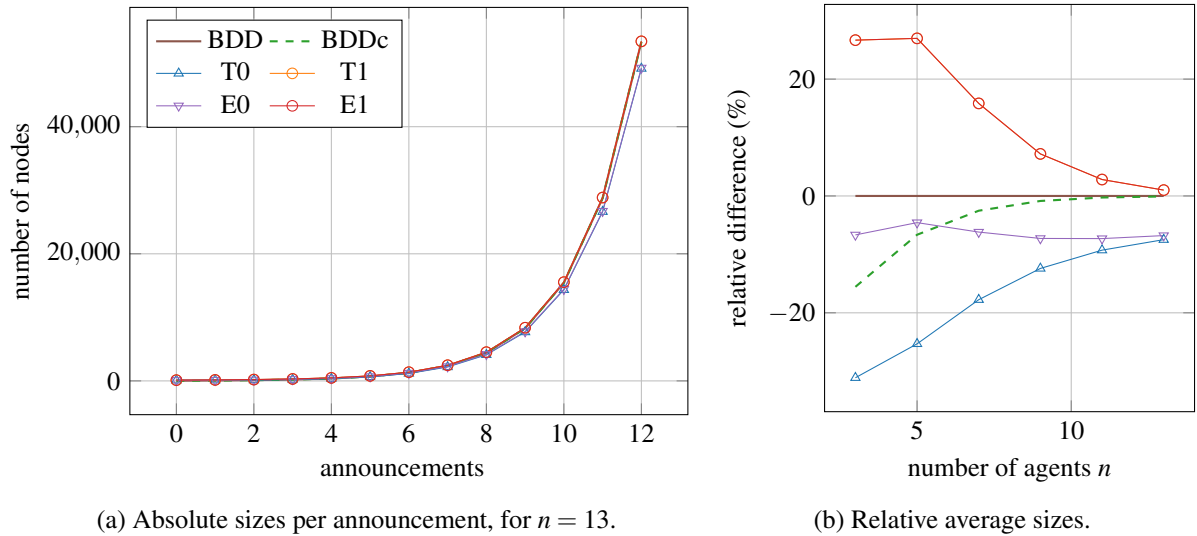

\paragraph{Dining cryptographers}
For $13$ agents we show the sizes after each announcement in Figure~\ref{fig:dining-resultplot}.
It becomes clear that there is little difference between the variants, which can be explained by the sparsity of the model throughout the whole story.
Still, the $T0$/$E0$ variants slightly outperform the BDD{(c)} and the $T1$/$E1$ variants.
This makes sense as most variables saying that agent $i$ paid will be false.
For lower numbers of agents the difference is larger, as visible in Figure~\ref{fig:dining-relative} where we vary the number of agents from $3$ to $13$.
Note that $T1$ and $E1$ overlap here, and $T0$ provides the best advantage.

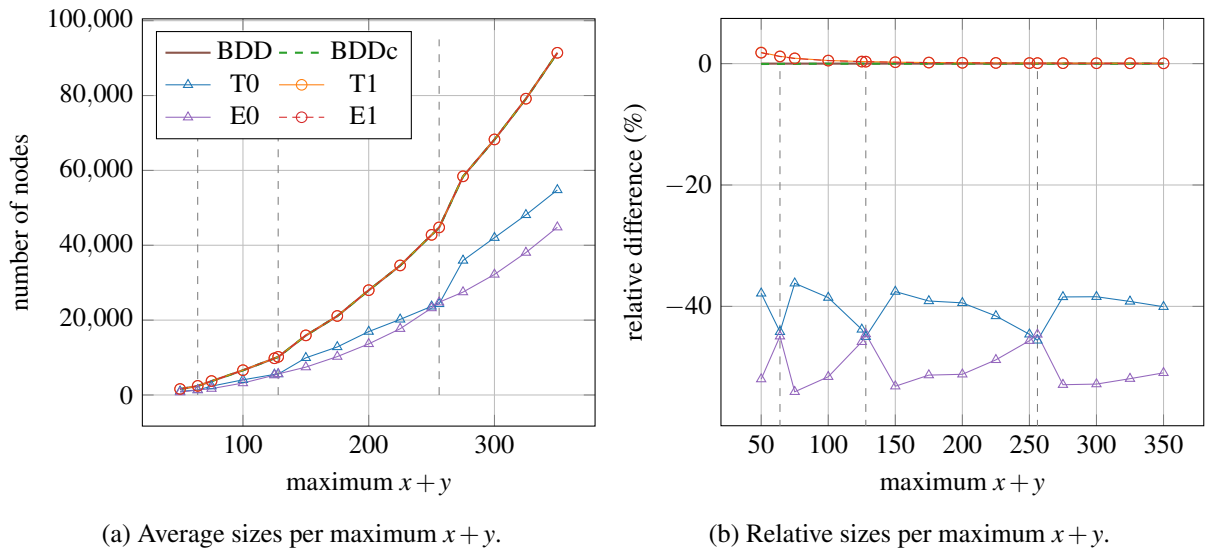
\begin{figure}[H]
  \centering
  \small
  \begin{subfigure}[b]{0.5\textwidth}
    \begin{tikzpicture}
      \pgfplotstableread{experimentResults/sap.dat}\sap
      \begin{axis}[legend columns=2, legend pos=north west, height=7cm, width=7.6cm, grid=major,
        scaled ticks=false, tick label style={/pgf/number format/fixed},
        ylabel=number of nodes,
        xlabel=maximum $x+y$,
        restrict expr to domain={\thisrow{round}}{-1:-1},
        filter discard warning=false, unbounded coords=discard]
        \addlegendentry{BDD}; \addplot+ [color=BDD,  mark=none, thick] table[x=n, y=BDD] {\sap};
        \addlegendentry{BDDc};\addplot+ [color=BDDc, mark=none, thick, dashed] table[x=n, y=BDDc] {\sap};
        \addlegendentry{T0};  \addplot+ [color=T0, mark=triangle] table[x=n, y=T0] {\sap};
        \addlegendentry{T1};  \addplot+ [color=T1, mark=o] table[x=n, y=T1] {\sap};
        \addlegendentry{E0};  \addplot+ [color=E0, mark=triangle] table[x=n, y=E0] {\sap};
        \addlegendentry{E1};  \addplot+ [color=E1, mark=o] table[x=n, y=E1] {\sap};
        \draw [dashed,gray] (axis cs:64,0) to (axis cs:64,95000);
        \draw [dashed,gray] (axis cs:128,0) to (axis cs:128,95000);
        \draw [dashed,gray] (axis cs:256,0) to (axis cs:256,95000);
      \end{axis}
    \end{tikzpicture}
    \caption{Average sizes per maximum $x + y$.}\label{fig:SAP-resultplot}
  \end{subfigure}
  \hfill
  \begin{subfigure}[b]{0.49\textwidth}
    \begin{tikzpicture}
      \pgfplotstableread{experimentResults/sap.dat}\sap
      \begin{axis}[height=7cm, width=8cm, grid=major,
        ylabel=relative difference (\%),
        xlabel=maximum $x+y$,
        restrict expr to domain={\thisrow{round}}{-1:-1},
        filter discard warning=false, unbounded coords=discard]
        \addplot+ [color=BDD,  mark=none, thick] table[x=n, y expr=0] {\sap};
        \addplot+ [color=BDDc, mark=none, thick, dashed] table[x=n, y expr=(\thisrow{BDDc}-\thisrow{BDD}) / \thisrow{BDD} * 100] {\sap};
        \addplot+ [color=T0, mark=triangle] table[x=n, y expr=(\thisrow{T0}-\thisrow{BDD}) / \thisrow{BDD} * 100] {\sap};
        \addplot+ [color=T1, mark=o] table[x=n, y expr=(\thisrow{T1}-\thisrow{BDD}) / \thisrow{BDD} * 100] {\sap};
        \addplot+ [color=E0, mark=triangle] table[x=n, y expr=(\thisrow{E0}-\thisrow{BDD}) / \thisrow{BDD} * 100] {\sap};
        \addplot+ [color=E1, mark=o] table[x=n, y expr=(\thisrow{E1}-\thisrow{BDD}) / \thisrow{BDD} * 100] {\sap};
        \draw [dashed,gray] (axis cs:64,-60) to (axis cs:64,0);
        \draw [dashed,gray] (axis cs:128,-60) to (axis cs:128,0);
        \draw [dashed,gray] (axis cs:256,-60) to (axis cs:256,0);
      \end{axis}
    \end{tikzpicture}
    \caption{Relative sizes per maximum $x + y$.}\label{fig:SAP-resultplot-relative}
  \end{subfigure}
  \caption{Results for Sum and Product.}
\end{figure}

\paragraph{Sum and Product}
In this last example we can vary the upper bound of $x+y$ from $50$ to $350$, but not the number of agents and announcements.
Figure~\ref{fig:SAP-resultplot} shows the sizes averaged over all four stages.
We note that the BDD{(c)}, $T1$ and $E1$ lines all overlap (with insignificant differences), and that T0 and E0 perform the best here.
In contrast to the first two examples, this advantage does not disappear for larger instances of the puzzle, as can be seen in Figure~\ref{fig:SAP-resultplot-relative} where we show the relative differences.
Interestingly, we see that $T0$ and $E0$ meet up and diverge again wherever the bound for $x+y$ is a power of 2 (i.e.\ 64, 128 or 256) which we mark by vertical dashed lines.
This is due to the bit-wise encoding where just above powers of two an additional bit is needed, but it must be false for almost all values.

\section{Conclusion}\label{sec:conclusion}

In all experiments we find a ZDD elimination rule that can reduce the number of nodes compared to BDDs, with the exception that in the Muddy Children example complement edges provide the same advantage.
This leads us to conclude that ZDDs are a promising tool for DEL model checking.
Specifically, if domain knowledge about the particular model allows one to predict which ZDD variant will be more compact, ZDDs can outcompete BDDs.

The BDD elimination rule treats true and false atomic propositions symmetrically, whereas ZDD rules are asymmetric.
This means their success depends on asymmetry in the model.

When translating an example from natural language to a formal models we usually try to avoid redundant variables, which already reduces the number of BDD-eliminable nodes.
This is likely the reason why using ZDDs provides an advantage or, for examples with a sparsity around 0.5 like the Muddy Children, at least the same performance as BDDs with complement edges.

Specifically for logic puzzles, usually all variables are needed, and models become asymmetric and sparse as information is revealed and possible answers are ruled out.
Our results confirm that sparsity and the kind of asymmetry prevalent in the model can predict which ZDD variant is most beneficial.

In this article we only considered S5.
SMCDEL also provides modules for K and in further experiments we compared the sizes of ZDDs and BDDs of the state law of \emph{belief} structures.
As an example we used the famous Sally-Anne false belief task.
The results were similar to those here and can be found in~\cite{MiedemaMSc2022}.

\paragraph{Future work}

An obvious limitation is that we only compared memory and not computation time.
The size of a decision diagram correlates with the computation time needed to build it.
But the step-wise construction techniques in SMCDEL are slower for ZDDs than for BDDs.
For example, to compute the Sum and Product result we rather convert each state law BDD to ZDDs instead of computing ZDDs directly.
Before a meaningful comparison of computation time can be done, the construction methods for ZDDs need to be further optimized.

We found some indicators which elimination rule is most compact in which case, but a more general approach to formalise domain knowledge and use it to make a correct prediction would be a powerful tool.

\paragraph{Acknowledgements}
This work is based on the master thesis~\cite{MiedemaMSc2022} by the first author, written at the University of Groningen and co-supervised by Rineke Verbrugge and the second author.

We thank the TARK reviewers for their careful reading and helpful comments on this article.

\bibliographystyle{eptcs}
\bibliography{literature}

\begin{thebibliography}{10}
\providecommand{\bibitemdeclare}[2]{}
\providecommand{\surnamestart}{}
\providecommand{\surnameend}{}
\providecommand{\urlprefix}{Available at }
\providecommand{\url}[1]{\texttt{#1}}
\providecommand{\href}[2]{\texttt{#2}}
\providecommand{\urlalt}[2]{\href{#1}{#2}}
\providecommand{\doi}[1]{doi:\urlalt{https://doi.org/#1}{#1}}
\providecommand{\eprint}[1]{arXiv:\urlalt{https://arxiv.org/abs/#1}{#1}}
\providecommand{\bibinfo}[2]{#2}

\bibitemdeclare{article}{vBvEGS2018:SMCDELbeyondS5}
\bibitem{vBvEGS2018:SMCDELbeyondS5}
\bibinfo{author}{Johan \surnamestart van Benthem\surnameend},
  \bibinfo{author}{Jan \surnamestart van Eijck\surnameend},
  \bibinfo{author}{Malvin \surnamestart Gattinger\surnameend} \&
  \bibinfo{author}{Kaile \surnamestart Su\surnameend} (\bibinfo{year}{2018}):
  \emph{\bibinfo{title}{Symbolic Model Checking for Dynamic Epistemic Logic —
  S5 and Beyond}}.
\newblock {\slshape \bibinfo{journal}{Logic and Computation}}
  \bibinfo{volume}{28}(\bibinfo{number}{2}), pp. \bibinfo{pages}{367--–402},
  \doi{10.1093/logcom/exx038}.

\bibitemdeclare{article}{bryant1986graph}
\bibitem{bryant1986graph}
\bibinfo{author}{Randal~E \surnamestart Bryant\surnameend}
  (\bibinfo{year}{1986}): \emph{\bibinfo{title}{Graph-based algorithms for
  boolean function manipulation}}.
\newblock {\slshape \bibinfo{journal}{IEEE Transactions on Computers}}
  \bibinfo{volume}{100}(\bibinfo{number}{8}), pp. \bibinfo{pages}{677--691},
  \doi{10.1109/TC.1986.1676819}.

\bibitemdeclare{incollection}{BryantBDDchapter}
\bibitem{BryantBDDchapter}
\bibinfo{author}{Randal~E. \surnamestart Bryant\surnameend}
  (\bibinfo{year}{2018}): \emph{\bibinfo{title}{Binary Decision Diagrams}}.
\newblock In \bibinfo{editor}{Edmund~M. \surnamestart Clarke\surnameend},
  \bibinfo{editor}{Thomas~A. \surnamestart Henzinger\surnameend},
  \bibinfo{editor}{Helmut \surnamestart Veith\surnameend} \&
  \bibinfo{editor}{Roderick \surnamestart Bloem\surnameend}, editors: {\slshape
  \bibinfo{booktitle}{Handbook of Model Checking}},
  \bibinfo{publisher}{Springer}, pp. \bibinfo{pages}{191--217},
  \doi{10.1007/978-3-319-10575-8_7}.

\bibitemdeclare{article}{burch1992symbolic}
\bibitem{burch1992symbolic}
\bibinfo{author}{Jerry~R \surnamestart Burch\surnameend},
  \bibinfo{author}{Edmund~M \surnamestart Clarke\surnameend},
  \bibinfo{author}{Kenneth~L \surnamestart McMillan\surnameend},
  \bibinfo{author}{David~L \surnamestart Dill\surnameend} \&
  \bibinfo{author}{Lain-Jinn \surnamestart Hwang\surnameend}
  (\bibinfo{year}{1992}): \emph{\bibinfo{title}{Symbolic model checking:
  $10^{20}$ states and beyond}}.
\newblock {\slshape \bibinfo{journal}{Information and computation}}
  \bibinfo{volume}{98}(\bibinfo{number}{2}), pp. \bibinfo{pages}{142--170},
  \doi{10.1016/0890-5401(92)90017-a}.

\bibitemdeclare{inproceedings}{charrierHintikkas2019}
\bibitem{charrierHintikkas2019}
\bibinfo{author}{Tristan \surnamestart Charrier\surnameend},
  \bibinfo{author}{Sébastien \surnamestart Gamblin\surnameend},
  \bibinfo{author}{Alexandre \surnamestart Niveau\surnameend} \&
  \bibinfo{author}{François \surnamestart Schwarzentruber\surnameend}
  (\bibinfo{year}{2019}): \emph{\bibinfo{title}{Hintikka's World: Scalable
  Higher-order Knowledge}}.
\newblock In: {\slshape \bibinfo{booktitle}{IJCAI 2019}}, pp.
  \bibinfo{pages}{6494--6496}, \doi{10.24963/ijcai.2019/934}.

\bibitemdeclare{article}{charrierSymbolic2019}
\bibitem{charrierSymbolic2019}
\bibinfo{author}{Tristan \surnamestart Charrier\surnameend},
  \bibinfo{author}{Sophie \surnamestart Pinchinat\surnameend} \&
  \bibinfo{author}{François \surnamestart Schwarzentruber\surnameend}
  (\bibinfo{year}{2019}): \emph{\bibinfo{title}{Symbolic model checking of
  public announcement protocols}}.
\newblock {\slshape \bibinfo{journal}{Logic and Computation}}
  \bibinfo{volume}{29}(\bibinfo{number}{8}), pp. \bibinfo{pages}{1211--1249},
  \doi{10.1093/logcom/exz023}.

\bibitemdeclare{article}{chaum1988dining}
\bibitem{chaum1988dining}
\bibinfo{author}{David \surnamestart Chaum\surnameend} (\bibinfo{year}{1988}):
  \emph{\bibinfo{title}{The dining cryptographers problem: Unconditional sender
  and recipient untraceability}}.
\newblock {\slshape \bibinfo{journal}{Journal of cryptology}}
  \bibinfo{volume}{1}(\bibinfo{number}{1}), pp. \bibinfo{pages}{65--75},
  \doi{10.1007/BF00206326}.

\bibitemdeclare{book}{van2007dynamic}
\bibitem{van2007dynamic}
\bibinfo{author}{Hans \surnamestart van Ditmarsch\surnameend},
  \bibinfo{author}{Wiebe \surnamestart van Der~Hoek\surnameend} \&
  \bibinfo{author}{Barteld \surnamestart Kooi\surnameend}
  (\bibinfo{year}{2007}): \emph{\bibinfo{title}{Dynamic Epistemic Logic}}.
\newblock \bibinfo{publisher}{Springer}, \doi{10.1007/978-1-4020-5839-4}.

\bibitemdeclare{article}{van2009publieke}
\bibitem{van2009publieke}
\bibinfo{author}{Hans \surnamestart van Ditmarsch\surnameend},
  \bibinfo{author}{Jan \surnamestart van Eijck\surnameend} \&
  \bibinfo{author}{Rineke \surnamestart Verbrugge\surnameend}
  (\bibinfo{year}{2009}): \emph{\bibinfo{title}{Publieke
  werken—freudenthal’s som-en-productraadsel}}.
\newblock {\slshape \bibinfo{journal}{Nieuw Archief voor Wiskunde}}
  \bibinfo{volume}{10}(\bibinfo{number}{2}), pp. \bibinfo{pages}{126--131}.
\newblock
  \urlprefix\url{https://www.nieuwarchief.nl/serie5/pdf/naw5-2009-10-2-126.pdf}.

\bibitemdeclare{article}{vDRv2007::SAPinDEL}
\bibitem{vDRv2007::SAPinDEL}
\bibinfo{author}{Hans \surnamestart van Ditmarsch\surnameend},
  \bibinfo{author}{Ji~\surnamestart Ruan\surnameend} \& \bibinfo{author}{Rineke
  \surnamestart Verbrugge\surnameend} (\bibinfo{year}{2007}):
  \emph{\bibinfo{title}{{Sum and Product in Dynamic Epistemic Logic}}}.
\newblock {\slshape \bibinfo{journal}{Logic and Computation}}
  \bibinfo{volume}{18}(\bibinfo{number}{4}), pp. \bibinfo{pages}{563--588},
  \doi{10.1093/logcom/exm081}.

\bibitemdeclare{article}{van2007epistemic}
\bibitem{van2007epistemic}
\bibinfo{author}{Jan \surnamestart van Eijck\surnameend} \&
  \bibinfo{author}{Simona \surnamestart Orzan\surnameend}
  (\bibinfo{year}{2007}): \emph{\bibinfo{title}{Epistemic verification of
  anonymity}}.
\newblock {\slshape \bibinfo{journal}{Electronic Notes in Theoretical Computer
  Science}} \bibinfo{volume}{168}, pp. \bibinfo{pages}{159--174},
  \doi{10.1016/j.entcs.2006.08.026}.

\bibitemdeclare{inproceedings}{gamblinSymbolic2022}
\bibitem{gamblinSymbolic2022}
\bibinfo{author}{Sébastien \surnamestart Gamblin\surnameend},
  \bibinfo{author}{Alexandre \surnamestart Niveau\surnameend} \&
  \bibinfo{author}{Maroua \surnamestart Bouzid\surnameend}
  (\bibinfo{year}{2022}): \emph{\bibinfo{title}{A Symbolic Representation for
  Probabilistic Dynamic Epistemic Logic}}.
\newblock In: {\slshape \bibinfo{booktitle}{AAMAS 2022}}, pp.
  \bibinfo{pages}{445--453}.
\newblock \urlprefix\url{https://dl.acm.org/doi/abs/10.5555/3535850.3535901}.

\bibitemdeclare{book}{knuth2011art4Ap1}
\bibitem{knuth2011art4Ap1}
\bibinfo{author}{Donald~E. \surnamestart Knuth\surnameend}
  (\bibinfo{year}{2011}): \emph{\bibinfo{title}{The Art of Computer
  Programming, volume 4A: Combinatorial Algorithms, Part 1}}.
\newblock \bibinfo{publisher}{Addison-Wesley}.

\bibitemdeclare{book}{littlewood1953methuen}
\bibitem{littlewood1953methuen}
\bibinfo{author}{John~E \surnamestart Littlewood\surnameend}
  (\bibinfo{year}{1953}): \emph{\bibinfo{title}{A Mathematician’s
  Miscellany}}.
\newblock \bibinfo{publisher}{Methuen and Company Limited}.

\bibitemdeclare{inproceedings}{lv2013cacbdd}
\bibitem{lv2013cacbdd}
\bibinfo{author}{Guanfeng \surnamestart Lv\surnameend}, \bibinfo{author}{Kaile
  \surnamestart Su\surnameend} \& \bibinfo{author}{Yanyan \surnamestart
  Xu\surnameend} (\bibinfo{year}{2013}): \emph{\bibinfo{title}{CacBDD: A BDD
  package with dynamic cache management}}.
\newblock In: {\slshape \bibinfo{booktitle}{Computer Aided Verification}},
  \bibinfo{organization}{Springer}, pp. \bibinfo{pages}{229--234},
  \doi{10.1007/978-3-642-39799-8_15}.

\bibitemdeclare{book}{mcmillan1993symbolic}
\bibitem{mcmillan1993symbolic}
\bibinfo{author}{Kenneth~L \surnamestart McMillan\surnameend}
  (\bibinfo{year}{1993}): \emph{\bibinfo{title}{Symbolic model checking}}.
\newblock \bibinfo{publisher}{Springer}, \doi{10.1007/978-1-4615-3190-6}.

\bibitemdeclare{mastersthesis}{MiedemaMSc2022}
\bibitem{MiedemaMSc2022}
\bibinfo{author}{Daniel \surnamestart Miedema\surnameend}
  (\bibinfo{year}{2022}): \emph{\bibinfo{title}{Zero-suppression Decision
  Diagrams versus Binary Decision Diagrams on Dynamic Epistemic Logic Model
  Checking}}.
\newblock Master's thesis, \bibinfo{school}{University of Groningen}.
\newblock \urlprefix\url{https://fse.studenttheses.ub.rug.nl/27287/}.

\bibitemdeclare{inproceedings}{minato1993zero}
\bibitem{minato1993zero}
\bibinfo{author}{Shin-ichi \surnamestart Minato\surnameend}
  (\bibinfo{year}{1993}): \emph{\bibinfo{title}{Zero-suppressed BDDs for set
  manipulation in combinatorial problems}}.
\newblock In: {\slshape \bibinfo{booktitle}{Proceedings of the 30th
  international Design Automation Conference}}, pp. \bibinfo{pages}{272--277},
  \doi{10.1145/157485.164890}.

\bibitemdeclare{article}{minato2001zero}
\bibitem{minato2001zero}
\bibinfo{author}{Shin-ichi \surnamestart Minato\surnameend}
  (\bibinfo{year}{2001}): \emph{\bibinfo{title}{Zero-suppressed BDDs and their
  applications}}.
\newblock {\slshape \bibinfo{journal}{International Journal on Software Tools
  for Technology Transfer}} \bibinfo{volume}{3}(\bibinfo{number}{2}), pp.
  \bibinfo{pages}{156--170}, \doi{10.1007/s100090100038}.

\bibitemdeclare{misc}{somenzi2004cudd}
\bibitem{somenzi2004cudd}
\bibinfo{author}{Fabio \surnamestart Somenzi\surnameend}
  (\bibinfo{year}{2012}): \emph{\bibinfo{title}{CUDD: CU decision diagram
  package}}.
\newblock \urlprefix\url{http://vlsi.colorado.edu/~fabio/CUDD/}.
\newblock \bibinfo{note}{Version 2.5.0}.

\bibitemdeclare{article}{SuVarForget2009}
\bibitem{SuVarForget2009}
\bibinfo{author}{K.~\surnamestart Su\surnameend},
  \bibinfo{author}{A.~\surnamestart Sattar\surnameend},
  \bibinfo{author}{G.~\surnamestart Lv\surnameend} \&
  \bibinfo{author}{Y.~\surnamestart Zhang\surnameend} (\bibinfo{year}{2009}):
  \emph{\bibinfo{title}{Variable Forgetting in Reasoning about Knowledge}}.
\newblock {\slshape \bibinfo{journal}{Journal of Artificial Intelligence
  Research}} \bibinfo{volume}{35}, pp. \bibinfo{pages}{677--716},
  \doi{10.1613/jair.2750}.

\end{thebibliography}

\section*{Appendix}

The ZDD encoding of knowledge structures has been integrated into SMCDEL itself.
All our results can be reproduced using the Haskell Tool \emph{Stack} from \url{https://haskellstack.org} as follows.

\begin{verbatim}
git clone https://github.com/jrclogic/SMCDEL
cd SMCDEL
git checkout zdd-experiments
stack bench --no-run-benchmarks # build but do not run yet
stack bench smcdel:bench:sizes-muddychildren
stack bench smcdel:bench:sizes-diningcryptographers
stack bench smcdel:bench:sizes-sumandproduct
\end{verbatim}

The last three commands will create \texttt{.dat} files containing the results.
On a system with a 4.8~GHz CPU the last three commands above take approximately 10 seconds, one minute and three hours.

We include the results for Dining Crytographers and Sum and Product here, but omit the (several pages long) results for the Muddy Children.

\paragraph{Results for Dining Cryptographers}
\phantom{.}

\begin{lstlisting}
# Note: round -1 indicates the average.
n	m	round	BDD	BDDc	T0	T1	E0	E1
3	1	0	9	7	9	13	11	13
3	1	1	15	12	10	19	14	19
3	1	2	21	19	12	25	17	25
3	1	-1	11.25	9.5	7.75	14.25	10.5	14.25
5	1	0	13	11	18	26	22	26
5	1	1	25	20	21	38	29	38
5	1	2	41	37	29	54	40	54
5	1	3	64	61	44	77	57	77
5	1	4	98	96	68	111	82	111
5	1	-1	60.25	56.25	45.0	76.5	57.5	76.5
7	1	0	17	15	31	43	37	43
7	1	1	35	28	36	61	48	61
7	1	2	61	55	50	87	67	87
7	1	3	102	97	79	128	100	128
7	1	4	170	166	133	196	157	196
7	1	5	285	282	228	311	254	311
7	1	6	479	477	388	505	415	505
7	1	-1	287.25	280.0	236.25	332.75	269.5	332.75
9	1	0	21	19	48	64	56	64
9	1	1	45	36	55	88	71	88
9	1	2	81	73	75	124	98	124
9	1	3	140	133	118	183	147	183
9	1	4	242	236	202	285	236	285
9	1	5	423	418	359	466	397	466
9	1	6	747	743	645	790	686	790
9	1	7	1326	1323	1156	1369	1199	1369
9	1	8	2352	2350	2052	2395	2096	2395
9	1	-1	1344.25	1332.75	1177.5	1441.0	1246.5	1441.0
11	1	0	25	23	69	89	79	89
11	1	1	55	44	78	119	98	119
11	1	2	101	91	104	165	133	165
11	1	3	178	169	161	242	198	242
11	1	4	314	306	275	378	319	378
11	1	5	561	554	494	625	544	625
11	1	6	1015	1009	906	1079	961	1079
11	1	7	1852	1847	1671	1916	1730	1916
11	1	8	3392	3388	3077	3456	3139	3456
11	1	9	6211	6208	5636	6275	5700	6275
11	1	10	11333	11331	10244	11397	10309	11397
11	1	-1	6259.25	6242.5	5678.75	6435.25	5802.5	6435.25
13	1	0	29	27	94	118	106	118
13	1	1	65	52	105	154	129	154
13	1	2	121	109	137	210	172	210
13	1	3	216	205	208	305	253	305
13	1	4	386	376	352	475	406	475
13	1	5	699	690	633	788	695	788
13	1	6	1283	1275	1171	1372	1240	1372
13	1	7	2378	2371	2190	2467	2265	2467
13	1	8	4432	4426	4106	4521	4186	4521
13	1	9	8277	8272	7687	8366	7771	8366
13	1	10	15449	15445	14341	15538	14428	15538
13	1	11	28764	28761	26628	28853	26717	28853
13	1	12	53342	53340	49156	53431	49246	53431
13	1	-1	28860.25	28837.25	26702.0	29149.5	26903.5	29149.5
\end{lstlisting}

\paragraph{Results for Sum and Product}
\phantom{.}

\begin{lstlisting}
# Note: round -1 indicates the average.
n	round	BDD	BDDc	T0	T1	E0	E1
50	0	5032	5030	3082	5060	2505	5060
50	1	697	696	458	726	288	724
50	2	547	546	361	576	221	574
50	3	1	1	0	31	0	31
50	-1	1569.25	1568.25	975.25	1598.25	753.5	1597.25
64	0	7916	7914	4247	7944	4602	7944
64	1	930	929	614	959	381	957
64	2	760	759	501	789	308	787
64	3	1	1	0	31	0	31
64	-1	2401.75	2400.75	1340.5	2430.75	1322.75	2429.75
75	0	12534	12532	7847	12566	5986	12566
75	1	1274	1273	900	1307	456	1305
75	2	939	938	658	972	335	970
75	3	35	34	27	68	12	66
75	-1	3695.5	3694.25	2358.0	3728.25	1697.25	3726.75
100	0	22438	22436	13514	22471	11279	22471
100	1	2289	2288	1567	2323	870	2321
100	2	1594	1593	1087	1628	596	1626
100	3	36	35	28	70	12	68
100	-1	6589.25	6588.0	4049.0	6623.0	3189.25	6621.5
125	0	33826	33824	18476	33859	19074	33859
125	1	3149	3148	2095	3183	1262	3181
125	2	2101	2100	1383	2135	835	2133
125	3	36	35	28	70	12	68
125	-1	9778.0	9776.75	5495.5	9811.75	5295.75	9810.25
128	0	35315	35313	18823	35348	20401	35348
128	1	3149	3148	2095	3183	1262	3181
128	2	2101	2100	1383	2135	835	2133
128	3	36	35	28	70	12	68
128	-1	10150.25	10149.0	5582.25	10184.0	5627.5	10182.5
150	0	55028	55026	33874	55065	26559	55065
150	1	5147	5146	3526	5185	1938	5183
150	2	3354	3353	2261	3392	1267	3390
150	3	40	39	32	78	12	76
150	-1	15892.25	15891.0	9923.25	15930.0	7444.0	15928.5
175	0	73233	73231	43763	73270	36869	73270
175	1	6753	6752	4635	6791	2549	6789
175	2	4265	4264	2893	4303	1599	4301
175	3	40	39	32	78	12	76
175	-1	21072.75	21071.5	12830.75	21110.5	10257.25	21109.0
200	0	98044	98042	58134	98082	49615	98082
200	1	8498	8497	5929	8537	3096	8535
200	2	5275	5274	3666	5314	1877	5312
200	3	41	40	33	80	12	78
200	-1	27964.5	27963.25	16940.5	28003.25	13650.0	28001.75
225	0	121863	121861	69555	121901	64632	121901
225	1	10103	10102	6910	10142	3827	10140
225	2	6284	6283	4289	6323	2317	6321
225	3	41	40	33	80	12	78
225	-1	34572.75	34571.5	20196.75	34611.5	17697.0	34610.0
250	0	148149	148147	80231	148187	83173	148187
250	1	14131	14130	8991	14170	6071	14168
250	2	8664	8663	5470	8703	3659	8701
250	3	41	40	33	80	12	78
250	-1	42746.25	42745.0	23681.25	42785.0	23228.75	42783.5
256	0	154815	154813	81895	154853	88925	154853
256	1	14992	14991	9616	15031	6371	15029
256	2	9084	9083	5787	9123	3786	9121
256	3	41	40	33	80	12	78
256	-1	44733.0	44731.75	24332.75	44771.75	24773.5	44770.25
275	0	203227	203225	123011	203269	98715	203269
275	1	18794	18793	12876	18837	7046	18835
275	2	11435	11434	7808	11478	4189	11476
275	3	45	44	37	88	12	86
275	-1	58375.25	58374.0	35933.0	58418.0	27490.5	58416.5
300	0	238864	238862	144850	238906	116069	238906
300	1	21339	21338	14568	21382	8066	21380
300	2	12671	12670	8634	12714	4662	12712
300	3	45	44	37	88	12	86
300	-1	68229.75	68228.5	42022.25	68272.5	32202.25	68271.0
325	0	277256	277254	165770	277298	137410	277298
325	1	24822	24821	16902	24865	9451	24863
325	2	14341	14340	9749	14384	5305	14382
325	3	45	44	37	88	12	86
325	-1	79116.0	79114.75	48114.5	79158.75	38044.5	79157.25
350	0	318340	318338	187632	318382	160813	318382
350	1	29838	29837	19932	29881	11776	29879
350	2	17313	17312	11500	17356	6686	17354
350	3	45	44	37	88	12	86
350	-1	91384.0	91382.75	54775.25	91426.75	44821.75	91425.25
\end{lstlisting}

\end{document}